\documentclass[twocolumn,prb,superscriptaddress]{revtex4}
\AtBeginDocument{
\heavyrulewidth=.08em
\lightrulewidth=.05em
\cmidrulewidth=.03em
\belowrulesep=.65ex
\belowbottomsep=0pt
\aboverulesep=.4ex
\abovetopsep=0pt
\cmidrulesep=\doublerulesep
\cmidrulekern=.5em
\defaultaddspace=.5em
}
\usepackage{amsmath,amssymb}
\usepackage{graphicx}
\usepackage[usenames,dvipsnames]{xcolor}
\usepackage{amsthm}
\usepackage{booktabs}
\usepackage{hyperref}
\usepackage[printwatermark]{xwatermark}
\usepackage[export]{adjustbox}
\usepackage[section]{placeins}
\usepackage[caption=false]{subfig}
\bibpunct{[}{]}{,}{n}{}{}
\definecolor{red1}{HTML}{FF4136}

\hypersetup{colorlinks=true,citecolor=Red,linkcolor=Blue,urlcolor=Black}
\usepackage[export]{adjustbox}

\begin{document}

\title{Pairing Tendencies in a Two-orbital Hubbard Model in One Dimension}

\author{N. D. Patel}
\affiliation{Department of Physics and Astronomy, The University of Tennessee, Knoxville, Tennessee 37996, USA}
\affiliation{Materials Science and Technology Division, Oak Ridge National Laboratory, Oak Ridge, Tennessee 37831, USA}

\author{A. Nocera}
\affiliation{Department of Physics and Astronomy, The University of Tennessee, Knoxville, Tennessee 37996, USA}
\affiliation{Materials Science and Technology Division, Oak Ridge National Laboratory, Oak Ridge, Tennessee 37831, USA}

\author{G. Alvarez}
\affiliation{Computer Science \& Mathematics %
Division and Center for Nanophase Materials Sciences, Oak Ridge National Laboratory, %
 \mbox{Oak Ridge, Tennessee 37831}, USA}

\author{A. Moreo}
\affiliation{Department of Physics and Astronomy, The University of Tennessee, Knoxville, Tennessee 37996, USA}
\affiliation{Materials Science and Technology Division, Oak Ridge National Laboratory, Oak Ridge, Tennessee 37831, USA}

\author{E. Dagotto}
\affiliation{Department of Physics and Astronomy, The University of Tennessee, Knoxville, Tennessee 37996, USA}
\affiliation{Materials Science and Technology Division, Oak Ridge National Laboratory, Oak Ridge, Tennessee 37831, USA}

\begin{abstract}
The recent discovery of superconductivity under high pressure 
in the ladder compound BaFe$_2$S$_3$ has opened a new field of research in iron-based superconductors
with focus on quasi one-dimensional geometries.
In this publication, using the Density Matrix Renormalization Group technique, 
we study a two-orbital Hubbard model defined in one dimensional chains. 
Our main result is the presence of hole binding tendencies at intermediate
Hubbard $U$ repulsion and robust Hund coupling $J_H/U=0.25$. Binding does not occur
neither in weak coupling nor at very strong coupling. The pair-pair correlations 
that are dominant near half-filling, or of similar strength as the charge and spin correlation channels, 
involve hole-pair operators that are spin singlets, use nearest-neighbor sites,
and employ different orbitals for each hole. The Hund coupling strength, presence of 
robust magnetic moments, and antiferromagnetic correlations among them are 
important for the binding tendencies found here. 
\end{abstract}

\maketitle

\section{Introduction}

High critical temperature superconductors 
based on iron represent one of the most important open problems in 
condensed matter physics~\cite{DCJohnston,Stewart,Peter,Scalapino,Chubukov,Dainature,RMP}.
Early considerations based on Fermi
surface nesting provided a robust starting point to 
rationalize their properties. However, the effect of repulsion between
electrons cannot be neglected~\cite{Dainature} as exemplified by the
existence of large magnetic moments at room temperature~\cite{moments1,moments2},
superconducting materials without hole pockets~\cite{nFSn}, as well as unexpectedly complex
spin arrangements~\cite{RMP}. Electronic correlation effects must be incorporated to
 understand the plethora of challenging results that experiments
are rapidly unveiling.

A new avenue of research for progress in the iron
superconductors family has recently opened. 
It has been shown experimentally that BaFe$_2$S$_3$~\cite{NatMatSC,ohgushi,HirataPRB15},
becomes superconducting at pressures above 10 GPa with an optimal critical temperature
$T_c = 24$~K. What is remarkable is that this material is not layered, like all other
iron-based superconductors, but instead has the geometry of a two-leg ladder. In other words,
they have a dominant crystal structure involving pairs of chains, ``legs'', that are coupled
via bonds of strength similar to those along the legs, dubbed the ``rungs''. 
This same compound, but at ambient pressure, is a Mott insulator with 
magnetic order involving ferromagnetic (FM) rungs and antiferromagnetic (AFM) 
legs and a N\'eel temperature $\sim 120$~K~\cite{NatMatSC}. 
Being the first iron-based superconductor that does not rely on layers, 
this discovery is conceptually exciting. From the theory perspective 
one-dimensional systems are often simpler than layers 
due to the availability of powerful computational techniques, thus robust 
many-body progress can be achieved in one dimension.

It is important to remark that there are other iron-based ladder materials 
with intriguing magnetic properties, although they have not been reported to be 
superconducting at high pressures yet. For example, the two-leg ladder BaFe$_2$Se$_3$
\cite{Feladder1,Feladder2,Feladder3,Feladder4,Feladder5,Feladder6,Feladder7,Feladder8,Feladder9,shuaiD}
is  an  insulator,  with  an activation energy 
0.13-0.178~eV~\cite{Feladder2,Feladder4}, 
long-range AFM order at $\sim$250~K, and robust low-temperature magnetic 
moments $\sim$2.8~$\mu_B$~\cite{Feladder1,Feladder2,Feladder3}. The dominant 
magnetic order at low temperature involves  2$\times$2 iron blocks with  
their  moments  aligned,  coupled  antiferromagnetically  
along the legs~\cite{Feladder1,Feladder4}. 
For the case of KFe$_2$Se$_3$, another two-leg ladder material, the
magnetic state is as in BaFe$_2$S$_3$ with FM rungs and AFM legs~\cite{Feladder5}. 
These same magnetic states were also found theoretically 
using the Hartree Fock approximation~\cite{Feladder7}. 
Their origin is non-trivial: the
2$\times$2 FM iron block patterns arise from frustration effects between fully 
FM tendencies at very large Hund coupling and AFM tendencies in all directions
at small Hund coupling~\cite{Feladder7}. Hartree-Fock results for 
layers~\cite{luo2D} and chains~\cite{luo1D} also revealed a similarly complex 
landscape of possible competing magnetic states once interactions are incorporated.

For proper context, it is necessary to recall that 
in the context of the copper-oxide high-$T_c$ superconductors, 
the theoretical and experimental study of two-leg ladder compounds 
made a considerable impact. In general, theorists can produce 
accurate results for quasi-one-dimensional systems and the
early predictions of subtle quantum effects, such as 
spin gaps and superconducting tendencies upon doping~\cite{ladder1,ladder2,ladder3},  
were later confirmed experimentally. For instance, high pressure experiments
for the two-leg ladder compound Sr$_{0.4}$Ca$_{13.6}$Cu$_{24}$O$_{41.84}$,
reported a superconducting critical temperature of 12~K~\cite{uehara}. 
A quantitative difference between Cu- and Fe-based ladders
is that the bridge between coppers is via an oxygen along the Cu-Cu bond, 
while in chalcogenides the bridge between irons is via chalcogen atoms 
located up and down the middle of the iron plaquettes. 
As a consequence, for the chalcogenides electronic
hoppings of similar strength are to be expected 
along legs, rungs, and also plaquette diagonals.

Although the computational study of two-leg ladder one-orbital Hubbard
and $t-J$ models were very successful in the context of the cuprates,
the case of the iron superconductors, even restricted to ladders, is more 
challenging. The reason is that multiorbital Hubbard models are needed
and even powerful techniques such as the Density Matrix Renormalization Group (DMRG)~\cite{SRWhite}
have difficulty in reaching sufficient accuracy for conclusive results.
In spite of these limitations, a recent publication~\cite{patel16} reported
progress in the study of a two-orbital model for BaFe$_2$S$_3$. In particular,
the magnetic order with FM rungs and AFM legs was qualitatively understood and 
clearly reproduced over robust portions of parameter space~\cite{aritaPRB15}. However,
the issue of pairing was more challenging due to severe size restrictions. In~\cite{patel16} 
it was assumed that high pressure causes doping of the two-leg ladders, a result
supported by recent density functional theory calculations~\cite{dong17}, 
and in agreement with
the Cu-oxide two-leg ladders context where experiments showed~\cite{piskunov} 
that indeed pressure transfers charge away from the ladders into chains effectively doping them.
Under these assumptions an intriguing result was unveiled: using 2$\times$8 clusters
indications of binding of two holes were observed at intermediate
values of the on-site Hubbard $U$ repulsion, and for a realistic Hund coupling $J_H/U = 0.25$. 
Note that this binding does not occur at small $U$, so in principle
it is outside the range of weak coupling expansions, and also does not occur in very strong
coupling~\cite{patel16}. Remarkably, the reported binding happens in a finite and intriguing 
range of $U/W$, where $W$ is the electronic bandwidth. However, the severe limitations
in size prevented us from reaching final conclusions in ~\cite{patel16} on whether the two-orbital
model on a two-leg ladder does superconduct or not.

To improve on the situation described above there are two avenues that we are simultaneously
pursuing. On one hand, the numerical aspects of the two-leg ladder model analysis described in the
previous paragraph must be substantially improved. Progress has been made and pairing on
2$\times$12 ladders has been confirmed recently (unpublished). Another avenue, pursued in
the present publication, is to search for models or geometries with similarities to those
of the real superconducting 
two-leg ladders but that would allow for the study of larger systems more comfortably. 

In this context, in the present publication using DMRG we study a two-orbital model 
that is mathematically 
similar to that used before for BaFe$_2$S$_3$, but now defined simply on a $chain$ 
as opposed to a two-leg ladder. When the binding energy is calculated vs. $U/W$ the
result, to be shown below, is similar to that found in the case of the two-leg ladder,
with binding observed at intermediate couplings and with a shape of the binding energy curve
vs. $U/W$ resembling that previously reported. Due to this similarity, it is reasonable
to believe that common physics causes the pairing tendencies both in ladders and in chains as
long as two orbitals are active. The
advantage of using chains, of course, is that much longer systems can be analyzed thus reducing
size effects. For all these reasons, in this publication a systematic study of a two-orbital
model defined on chains is presented, with emphasis on pairing and superconducting tendencies.
The analysis is presented in a systematic manner, varying the many 
couplings and electronic densities and even further boosting pairing tendencies by introducing
extra Heisenberg interactions. Overall, our analysis concludes that it is the Hund coupling $J_H$
that primarily drives the pairing tendencies, supplemented by AFM tendencies between the effective
$S=1$ spins of the undoped sites. As explained
below, it is known that there are unrealistic ranges of couplings where $J_H$
explicitly boosts hole attraction. What
is remarkable of our results described here is 
that \emph{similar tendencies survive into the realistic
regime $J_H/U=0.25$ where the model is not explicitly 
attractive} because of the competition between $J_H$
with the interorbital repulsion $U'$. These promising results 
are preliminary steps towards a clarification of the origin of pairing in iron-based superconductors,
but more work is needed to establish definitely that pairing of electronic origin
is active in two-orbital Hubbard models.

The organization of the manuscript is as follows. Section~II 
provides details of the model, technique used, and observables measured. 
Section~III contains the main results, addressing both magnetic and pairing properties
of the model under scrutiny. Section~IV contains our main discussion and conclusions.

\FloatBarrier
\section{Model and Method} \label{se:model}

The multiorbital Hubbard model used in this publication is defined as 
\begin{equation} \label{eq:Hubb}
\begin{split}
H &= -t \sum_{\substack{\langle i j \rangle \\ \gamma \sigma}} (c^{\dagger}_{i \gamma \sigma} 
c^{\phantom{\dagger}}_{j \gamma \sigma} + h.c.)
	 + U \sum_{i \gamma} n_{i \gamma \uparrow} n_{i \gamma \downarrow} \\
	& +	(U'- \frac{J_{H}}{2}) \sum_{\substack{i \\ \gamma < \gamma'}} n_{i \gamma} n_{i \gamma'} 
	 - 2J_{H} \sum_{\substack{i \\ \gamma < \gamma'}} {{\mathbf{S}_{i \gamma}}\cdot{\mathbf{S}_{i \gamma'}}} \\
	& + J_{H} \sum_{\substack{i \\ \gamma < \gamma'}} (P_{i\gamma}^{\dagger} P_{i\gamma'}^{\phantom{\dagger}} 
+ \text{h.c.}),
\end{split}
\end{equation}
where $c^{\dagger}_{i \gamma \sigma}$ ($c^{\phantom{\dagger}}_{i \gamma \sigma}$) creates (annihilates) 
an electron at site $i$ of a chain, with orbital $\gamma$ (either $a$ or $b$), and
spin projection along the $z$-axis $\sigma$. The first term represents the kinetic energy 
of the electrons. Note that for simplicity, the 2$\times$2 hopping matrix is the unit matrix
i.e. only hoppings between the same orbitals is allowed.  Although our overarching goal is
the understanding of iron-based superconductors, these hoppings do not intend to represent
the tunneling amplitudes of any particular material but they are chosen for simplicity.
The second is the standard on-site Hubbard repulsion $U$ between spins $\uparrow$ and
$\downarrow$ electrons. The third term is the repulsion between electrons at different 
orbitals. As shown in many previous
publications, besides the canonical $U'$ repulsion the coupling strength affecting this
term contains a contribution regulated by the Hund coupling $J_H$.
Fourth is the portion that explicitly shows the ferromagnetic character
of the Hund interaction. The last term is the pair hopping. The number operator is defined as
$n_{i\sigma\gamma} = c_{i\sigma\gamma}^{\dagger}c^{\phantom{\dagger}}_{i\sigma\gamma}$ 
and the pair as $P_{i \gamma} = c_{i \gamma \uparrow} c_{i \gamma
\downarrow}$. The standard relation $U' = U - 2J_{H}$ is assumed. 
While many of the results are for $J_{H}/U = 0.25$, considered realistic and used in the
previous publication for ladders~\cite{patel16}, in some of the results below the Hund coupling is varied.
The bandwidth corresponding to the kinetic energy portion is $W=4t$ and the Hubbard strength
will be provided primarily as $U/W$.
The hopping is the unit of energy $t=1.0$, unless stated otherwise. 

To obtain our results we use the DMRG technique with open boundary conditions 
with focus on the ground-state of the two-orbital chain, employing 
at least $1600$ states. Most of the results are for a $32$ sites two-orbital chain, 
unless stated otherwise, while some of the 
results were confirmed using up to $64$ sites. Truncation error 
remain below $\sim 10^{-6}$ for all of our results.

We have measured several observables. The
binding energy that is an indicator for pairing tendencies~\cite{RMP94} is defined as 
%
\begin{equation} \label{eq:BindEn}
\Delta E = E_{N} + E_{N-2} - 2 E_{N-1}, 
\end{equation}
%
where $E_{N}$, $E_{N-1}$, and $E_{N-2}$ are the total ground 
state energy of the half-filled, $1$-hole doped, and $2$-hole 
doped systems. Here, $N=2L$ with $L$ the length of the chain. 
The real space charge and spin correlations are 
%
\begin{equation} \label{eq:chargeR}
N(R) = \frac{1}{N_R} \sum_{|i-j| = R} [\langle n_i n_{j} \rangle - \langle n_i \rangle \langle n_{j} \rangle], 
\end{equation}
\begin{equation} \label{eq:spinR}
Sp(R) = \frac{1}{N_R} \sum_{|i-j| = R} \langle {{\mathbf{S}_i}\cdot{\mathbf{S}_{j}}} \rangle,
\end{equation}
%
where $N_{R}$ is the number of neighbours at distance $R$ 
from site $i$ (namely, averages over pairs of sites at equal distance
are performed). The Fourier transform of $Sp(R)$ is the spin structure factor 
$Sp(k)$

To study the effects of holes on the magnetic correlations,
we define a projector operator $\mathcal{P}_{i\gamma}$ that projects out
the portion of the ground state where site $i$ and orbital $\gamma$ are
occupied~\cite{projector}:
%
\begin{equation} \label{eq:Proj}
\mathcal{P}_{i \gamma} = c^{\phantom{\dagger}}_{i \gamma \uparrow} c^{\dagger}_{i \gamma \uparrow} 
c^{\phantom{\dagger}}_{i \gamma \downarrow} c^{\dagger}_{i \gamma \downarrow}.
\end{equation}
%
To work in the Hilbert space corresponding to $n_{h}$
number of holes at specific locations, we apply a product
of projectors onto the ground state with $n_{h}$ holes. 
For example, $\mathcal{P}_{6a}\mathcal{P}_{9b}|\psi_{N-2}\rangle$ projects out the
occupied part of the two-hole ground state on orbital $a$ at site $6$, and 
on orbital $b$ at site $9$. We also calculate the
local spin-spin correlations $ \langle \psi | {{\mathbf{S}_{i\gamma}}\cdot{\mathbf{S}_{j\gamma'}}} 
\mathcal{P} | \psi \rangle / \langle \psi |\mathcal{P} | \psi \rangle$,
where the maximum possible magnitude of the correlations is 3/4.

There are many possible superconducting
pair correlations that one can explore for this system. 
Due to the local 
inter- and intra-orbital Coulomb repulsion, on-site pairing is 
not expected to dominate~\cite{RMP94}.
Thus, pairing operators for 
two electrons at nearest-neighbor (NN) sites $i$ and $i+1$ will be considered in analogy with the approach taken in 
other purely electronic models where magnetic properties trigger pairing.

An intraorbital nearest neighbor pairing operators is
\begin{equation}\label{NNpairingop1}
{\Delta_{nn,-}^{\gamma\gamma}}^{\dagger}(i)= c^{\dagger}_{i,\gamma,\uparrow} c^{\dagger}_{i+1,\gamma,\downarrow}  - 
 															   c^{\dagger}_{i,\gamma,\downarrow} c^{\dagger}_{i+1,\gamma,\uparrow}, 
\end{equation}
\noindent which creates a pair of electrons at nearest neighboring sites $i$ and $i+1$ 
in orbital $\gamma$ and forming a spin singlet. The corresponding pairing operator for the case in 
which the two electrons form a spin triplet is given by
\begin{equation}\label{NNpairingop2}
{\Delta_{nn,+}^{\gamma\gamma}}^{\dagger}(i)= c^{\dagger}_{i,\gamma,\uparrow} c^{\dagger}_{i+1,\gamma,\downarrow}  + 
 															   c^{\dagger}_{i,\gamma,\downarrow} c^{\dagger}_{i+1,\gamma,\uparrow}.
\end{equation}

Since the Hamiltonian is invariant under an orbital exchange only orbital symmetric combinations of the intraorbital pairing correlations have to be considered. 
An interorbital nearest neighbor pairing operator is given by
\begin{equation}\label{NNpairingop3}
{\Delta_{nn,-}^{ab}}^{\dagger}(i)= c^{\dagger}_{i,a,\uparrow} c^{\dagger}_{i+1,b,\downarrow}  - 
 															   c^{\dagger}_{i,a,\downarrow} c^{\dagger}_{i+1,b,\uparrow}, 
\end{equation}
\noindent which creates two electrons at site $i$ and orbital $a$ and site $i+1$ and orbital $b$, 
forming a spin singlet (the orbital exchanged pairing operator is identical 
due to the orbital symmetry).
The interorbital pairing operator that creates the electrons in a triplet state is given by
\begin{equation}\label{NNpairingop4}
{\Delta_{nn,+}^{ab}}^{\dagger}(i)= c^{\dagger}_{i,a,\uparrow} c^{\dagger}_{i+1,b,\downarrow}  + 
 															   c^{\dagger}_{i,a,\downarrow} c^{\dagger}_{i+1,b,\uparrow}. 
\end{equation} 

The pair-pair correlations are given by
\begin{equation}\label{NNpairs2}
 \mathcal{O}^{\gamma\gamma'}_{nn,\pm}(R) = \frac{1}{2N_{R}} \sum_{i} \langle{\Delta_{nn,\pm}^{\gamma\gamma'}}^{\dagger}(i) \Delta_{nn,\pm}^{\gamma\gamma'}(i+R) \rangle,
\end{equation}
\noindent where $\pm$  indicates if the pair is a spin triplet or singlet, 
and $\gamma$ and $\gamma'$ indicate orbitals $a$ or $b$.  Since the results 
explicitly presented in this manuscript are for the interorbital
spin singlet $\mathcal{O}^{ab}_{nn,-}(R)$ and triplet 
correlations  $\mathcal{O}^{ab}_{nn,+}(R)$,
below we will use the notation $\mathcal{O}^{ab}_{nn,-}(R) \equiv S^{ab}_{nn}(R)$ 
and $\mathcal{O}^{ab}_{nn,+}(R) \equiv T^{ab}_{nn}(R)$, respectively. 
Analogous on-site pairing operators were also considered but
their correlations always decayed faster than the dominant NN sites pair-pair 
correlation ($S^{ab}_{nn}(R)$ as shown below). For this reason, the actual 
expressions for on-site operators are not provided explicitly.  

We have measured other observables as well. For example, by averaging
the pair correlations over a finite intermediate portion of the chain we can
reduce short distance effects, that sometimes lead to believe that pairing
is dominant even if the long distance tail is small, and also to reduce boundary effects
caused by the open boundary conditions. Here we define the pairing strength as 
\begin{equation} \label{eq:pstregth}
\bar{D} = \sum_{R=7}^{12} | S^{ab}_{nn}(R)|,
\end{equation}
where we have used the spin-singlet nearest-neighbor combination explicitly because it will be
shown below that it is dominant in our study.

%
\begin{figure}[thbp]
\begin{center}
 \includegraphics[trim = 0mm 0mm 0mm 0mm,
height=0.35\textwidth,width=0.38\textwidth,angle=0]{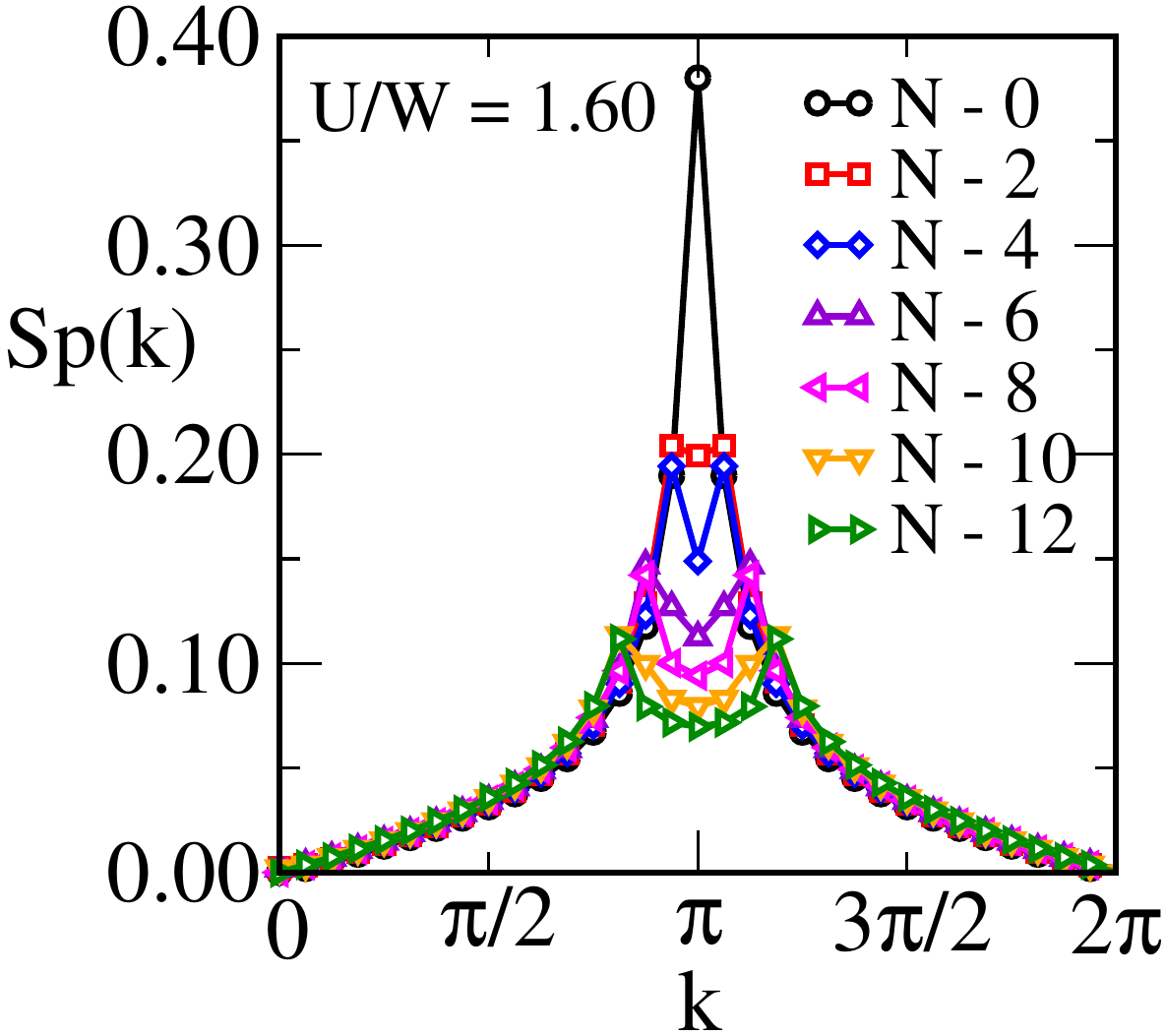}
\end{center}
\vspace{-0.6cm}
\caption{(color online) Spin structure factor vs. wavevector $k_x$  
along the chain direction. Results correspond to $U/W = 1.60$, $J_H/U=0.25$, 
various hole dopings as indicated, and employing a $32$ sites chain ($N=64$) and the DMRG technique.
In this figure, and others not shown at several values of $U/W$,  
the peak at $k_x = \pi$ denoting staggered order at half-filling 
becomes incommensurate upon hole doping.
}
\label{fig:Skx}
\end{figure}
%

\FloatBarrier
\section{Results}

In this section, the main results will be described. The 
language to be used below should always be considered in the framework of one
dimensional systems where long-range order is not possible. For example, expressions
such as ``staggered AFM order'' indicate that staggered spin arrangements decay the slowest
with distance as compared with other patterns, but eventually all correlation functions decay to
zero with increasing distance in one dimension with short-range interactions.

As expressed before, we remind the readers that experimentally in the Fe-ladders superconductivity appeared with increasing pressure, not with explicit hole doping. However, similarly as in the case of the Cu-ladders, it is believed that pressure may lead to a rearrangement of charge particularly with regards to the average number of electrons at the iron atoms. This perception is supported by recent $ab$-$initio$ calculations~\cite{dong17}. As a consequence, in our effort described below we will search for pairing indications by doping with holes the half-filled system, rather than modelling pressure directly.

\subsection{Magnetic order and local moments}

\begin{figure}[thbp]
\begin{center}
 \includegraphics[trim = 0mm 0mm 0mm 0mm,
height=0.23\textwidth,width=0.49\textwidth,angle=0]{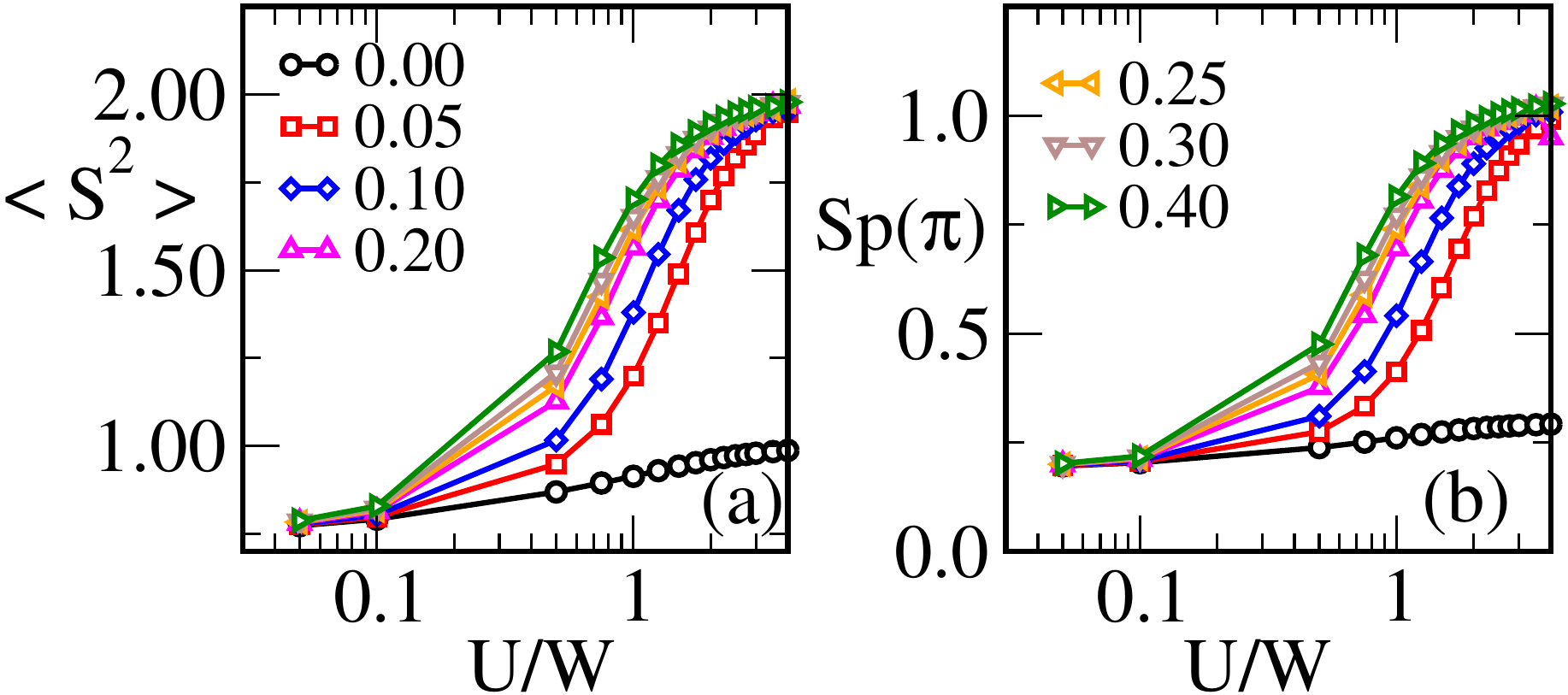}
\end{center}
\vspace{-0.6cm}
\caption{(color online) (a) Spin squared $\langle S^2 \rangle$ and (b)
staggered spin structure factor $Sp(\pi)$ vs. 
$U/W$ for a half-filled $8$-sites system using various Hund's coupling 
($J_H/U$) as indicated. At zero $J_H/U$, local moments are not developed up to large 
$U/W$ and therefore there is no robust magnetic ordering. 
}
\label{fig:fig5_NMag}
\end{figure}

Let us start our computational analysis of the two-orbital Hubbard model
defined in the previous Section by focusing on the magnetic order. 
Figure~\ref{fig:Skx} contains the spin structure factor at $U/W=1.60$, 
a coupling strength of much importance for pairing as shown below, for different
number of holes. For the case of half-filling, $N=64$ electrons for the 32-sites chain
of focus in Fig.~\ref{fig:Skx}, the spin order is clearly of the staggered
antiferromagnetic (AFM) form as expected. In this regime of Hubbard couplings the local spin
at every site is already well developed and close to the spin-1 limit, as shown in 
panel (a) of Fig.~\ref{fig:fig5_NMag} for most Hund couplings studied, with the exception
of $J_H = 0$. Thus, this AFM correlations are compatible with the
spin correlations of a Haldane spin-1 chain. Our study of a two-orbital Hubbard model, instead of
a Heisenberg model, involves energy scales much higher than those typical of the integer-spin
chains and for this reason we will not focus on subtle issues such as
spin-gaps in the system. 
Panel (b) of Fig.~\ref{fig:fig5_NMag} shows that together with the development of the
spin-1 moments at every site also robust AFM correlations develop at least at short distances, 
again with the exception of $J_H = 0$.

%
%

As the doping of holes increases, Fig.~\ref{fig:Skx} illustrates that spin
incommensurate (IC) correlations develop smoothly. While this spin IC order is compatible
with spin excitations from the band dispersion in the kinetic energy portion of the model,
note that $U/W = 1.60$ is already in intermediate coupling. Analysis of the spin-spin
correlations ``across holes''~\cite{martins1,martins2,martins3}, 
to be shown in more detail below, indicate that the spins tend to arrange and couple
in a manner qualitatively compatible with the exact results for the
one-orbital Hubbard model at $U=\infty$~\cite{ogata}.
This arrangement is the most optimal to
favor simultaneously the hole mobility and spin correlations, and it is
qualitatively different from the analysis based on 
Fermi surface characteristics.

%
\begin{figure}[thbp]
\begin{center}
 \includegraphics[trim = 0mm 0mm 0mm 0mm,
 height=0.34\textwidth,width=0.36\textwidth,angle=0]{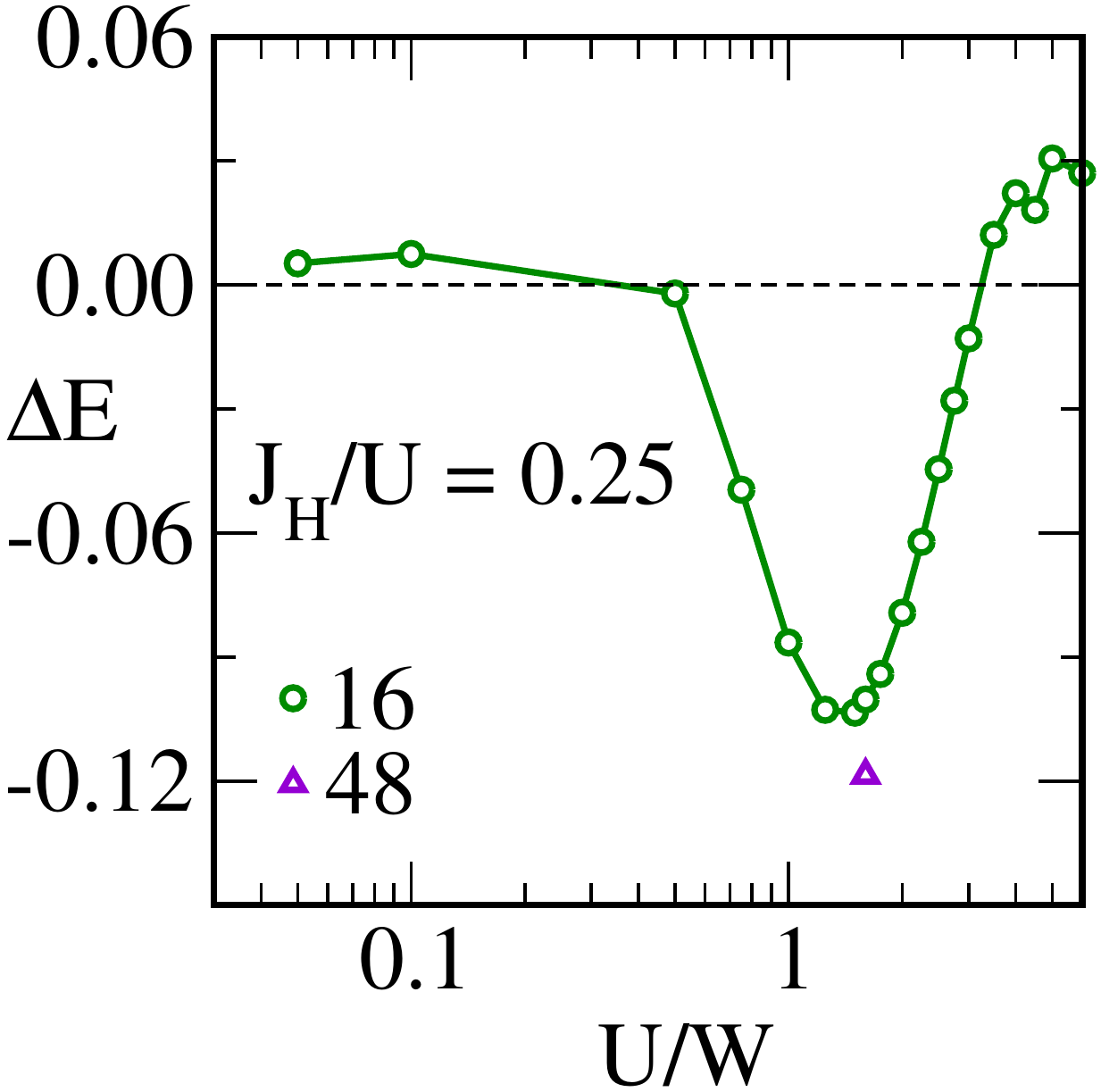}
\end{center}
\vspace{-0.6cm}
\caption{(color online)
Binding energy ($\Delta E$) vs. $U/W$ at a fixed
$J_H/U = 0.25$, and using a 16-sites chain. For intermediate 
interaction strength, there is a wide range with
negative binding energy indicating a region where holes pair. 
}
\label{fig:fig1a_BindingvsU}
\end{figure}

\subsection{Hole pairs and their internal structure}

\begin{figure}[thbp]
\begin{center}
 \includegraphics[trim = 0mm 0mm 0mm 0mm,
 height=0.34\textwidth,width=0.36\textwidth,angle=0]{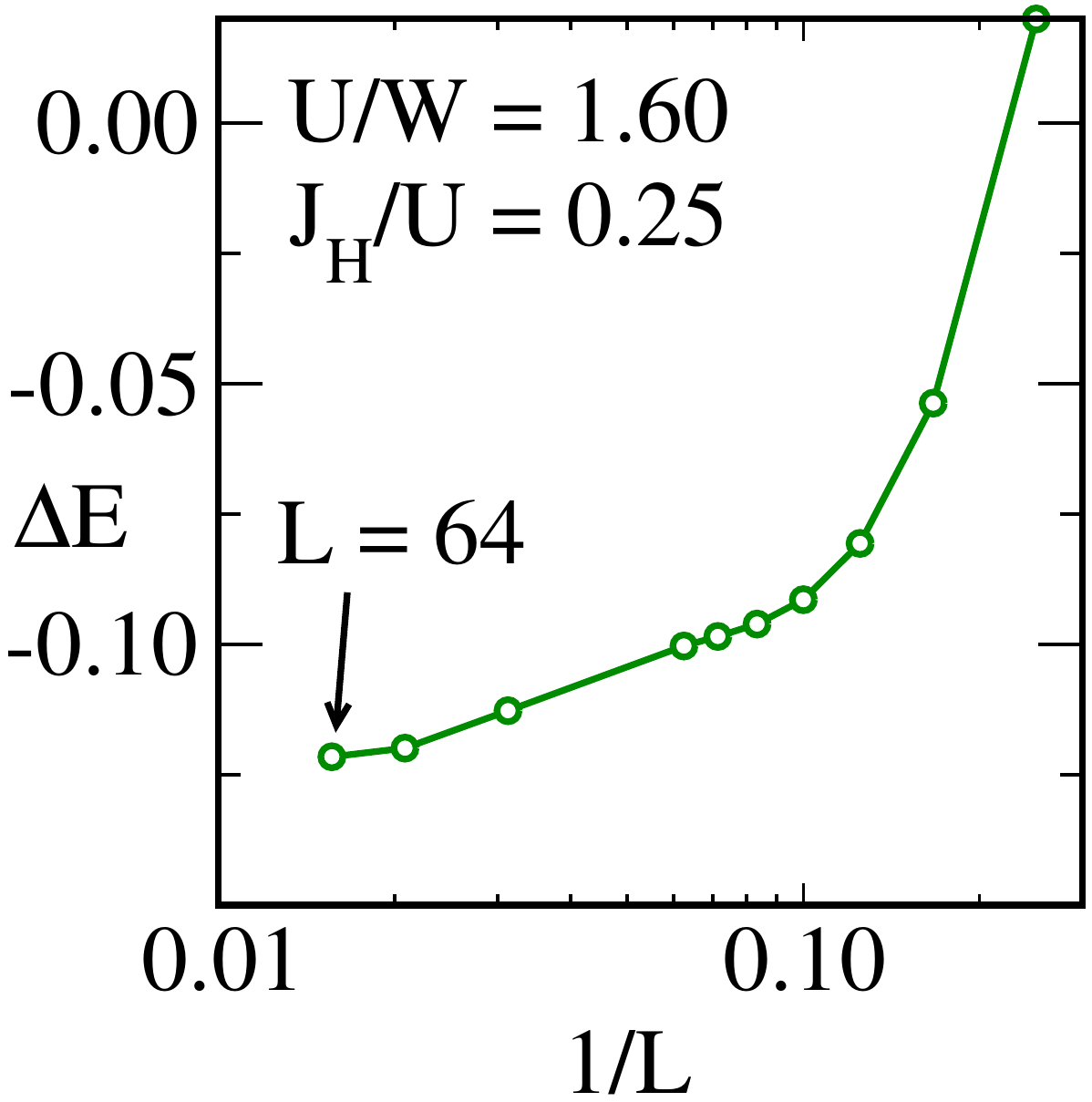}
\end{center}
\vspace{-0.6cm}
\caption{(color online)
Scaling of $\Delta E$ vs. inverse chain length (1/$L$) at coupling
$U/W = 1.60$ where there is robust negative binding. 
In the bulk limit, $\Delta E$ remains negative according to the extrapolation of these results.
}
\label{fig:fig1b_Bindingvs1OL}
\end{figure}

The main result of this publication is that the model studied here presents a regime of hole pair binding
that correlates with robust pair-pair correlations in a spin-singlet channel, 
as will be described below. Figure~\ref{fig:fig1a_BindingvsU} shows the binding energy $\Delta E$, as defined
in Eq.~\ref{eq:BindEn}, for the case of two holes added to half-filling, 
varying $U/W$ at a fixed Hund coupling $J_H/U=0.25$. Starting approximately
at $U/W \sim 0.6$ and up to $U/W \sim 3.0$, the binding energy is negative indicative of the formation
of a bound state of two holes. Considering recent developments in the study of iron-based
superconductors~\cite{Dainature} this regime of $U/W$ is realistic. Moreover, the
Hund coupling value is also in a reasonable range for pnictides and chalcogenides that are well known
for having a robust Hund-driven physics. The results in Fig.~\ref{fig:fig1a_BindingvsU} were obtained using
a 16-sites chain but they appear robust varying the length of the system. 
For instance, approximately at the minimum of the curve at $U/W = 1.60$ 
results for 48 sites are only slightly more negative than for 16 sites. 
Figure \ref{fig:fig1b_Bindingvs1OL} contains a size scaling analysis of binding at $U/W=1.60$
illustrating this conclusion. Our best efforts indicate 
that size effects are small and moreover with increasing
chain length the binding magnitude slightly increases in absolute value. 
Thus, the bulk-limit binding energy at $U/W=1.60$ appears to be close to -0.13 in hopping units.

%
%

%

Besides the surprising result that binding is possible even 
in the presence of a strong Hubbard $U$ repulsion, it is interesting to remark
the similarity of Fig.~\ref{fig:fig1a_BindingvsU} with the binding results found 
before in the context of two-leg ladders (see Fig.~8 of~\cite{patel16}).
In both cases, ladders (short sizes were studied in~\cite{patel16}) and chains, 
$\Delta E$ starts positive with increasing $U/W$,
drops to negative at intermediate couplings where it remains into the strong coupling
regime, and then it becomes positive again at abnormally large $U/W$. 
Note that the region where
$\Delta E$ is positive is not important: in the bulk limit the energy of two holes that do not
form a bound state should converge to the energy of two independent holes, rendering $\Delta E$
equal to zero. But the negative region of $\Delta E$ is physically realistic and representative
of pair formation: two holes lower the energy of the system by being close to each other.

%
\begin{figure}[thbp]
\begin{center}
 \includegraphics[trim = 0mm 0mm 0mm 0mm,
 height=0.30\textwidth,width=0.43\textwidth,angle=0]{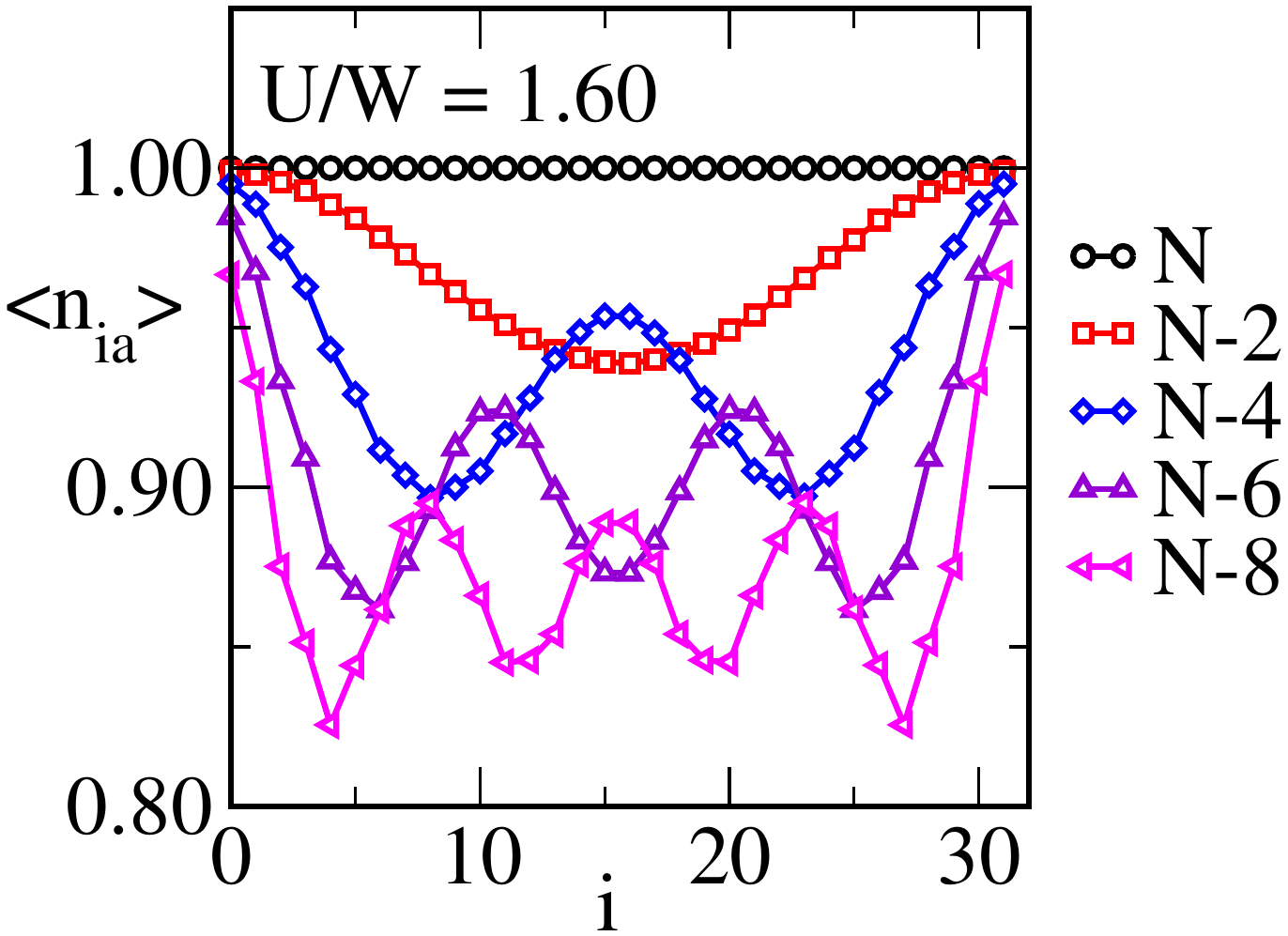}
\end{center}
\vspace{-0.6cm}
\caption{(color online) Local charge density of orbital $a$ at $U/W=1.60$, i.e. 
in the binding region, using a $32$-sites chain and  $J_H/U = 0.25$. 
For a fixed number of holes $n_h$, we find $n_h/2$ number of ``dips'' in
the charge density. Note that the charge density profile of orbital 
$b$ is the same as that of orbital $a$, by symmetry.
}
\label{fig:fig2_localoccupation}
\end{figure}
%

Figure ~\ref{fig:fig2_localoccupation} provides the electronic density using a 32-sites chain,
corresponding to the orbital $a$ (the results for $b$ are identical, because the model is
invariant if $a$ and $b$ are exchanged). For half-filling, the density is virtually 
equal to one at all sites. What is interesting is that for 2 holes,
there is only 1 minimum indicative of the existence of a hole pair. For 4 holes there are 2
minima, for 6 holes 3 minima, and for 8 holes 4 minima. All these results are 
at least compatible with the existence of hole pair formation,
as the binding energy indicates.

%
\begin{figure}[thbp]
\begin{center}
\includegraphics[trim = 0mm 0mm 0mm 0mm,
height=0.34\textwidth,width=0.37\textwidth,angle=0]{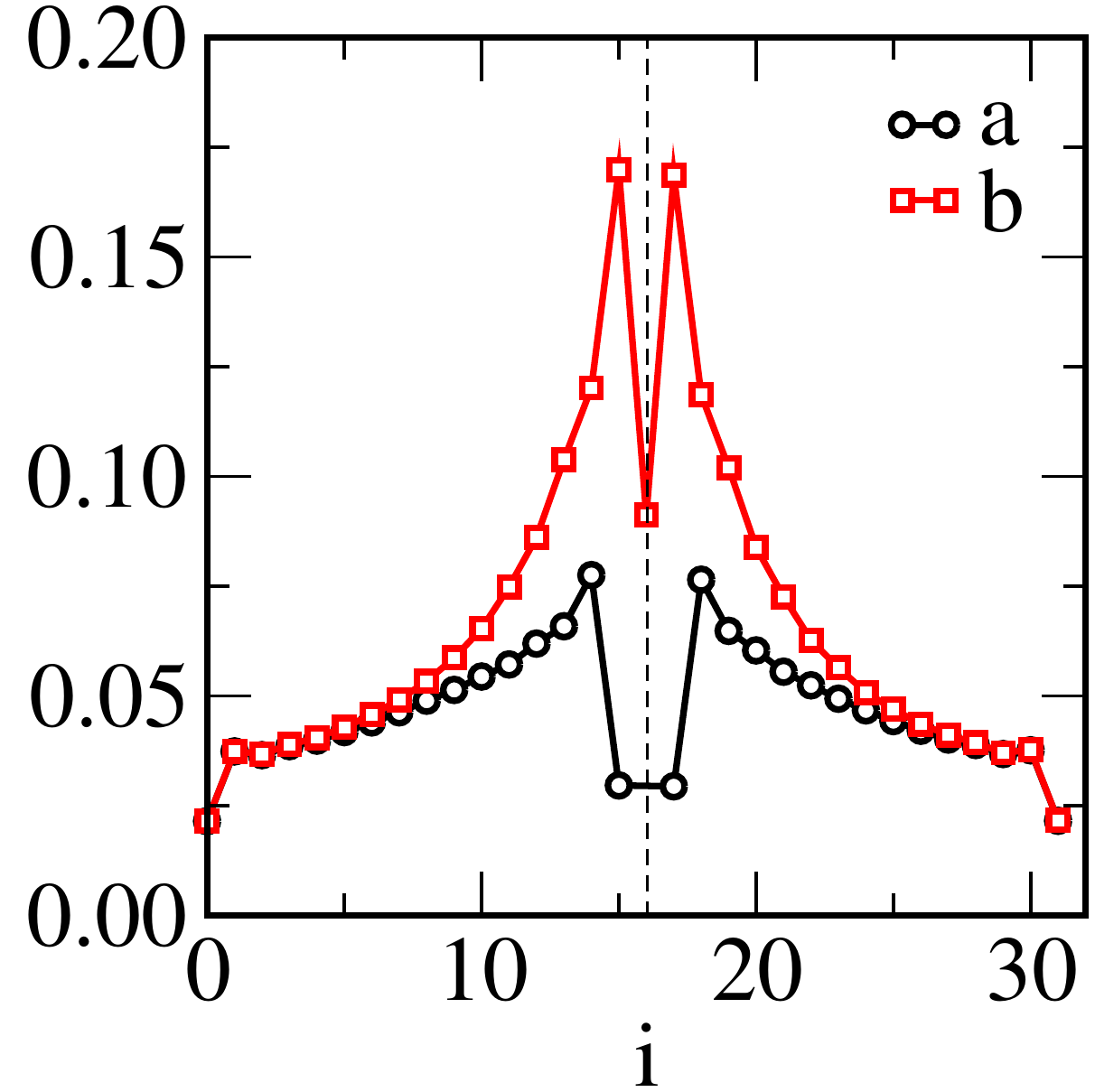}
\llap{\rotatebox{90}{{\parbox[a]{6.8cm}{\Large$\boldsymbol{\langle \mathcal{P}_{16a} 
\mathcal{P}_{i \gamma} \rangle}$\\\rule{0ex}{6.6cm}}}}}
\end{center}
\vspace{-0.6cm}
\caption{(color online) Probability of finding a hole at site $i$ and orbital $\gamma$
given that the other hole is projected from the two-holes ground state to be 
at site $16$, orbital $a$, of a two-orbital chain with 32
sites, at $U/W=1.60$ and $J_H/U=0.25$. The result 
is normalized to $\langle \mathcal{P}_{16a} \rangle$.
We find the largest probability
of the non-projected hole to be in the other orbital $b$ and 
at the neighboring site. In the ladder analogy of the two-orbital chain 
(see Fig.~\ref{fig:fig6_ProjSS}), these dominant pairs are equivalent 
to pairs along the diagonal of effective plaquettes. 
}
\label{fig:Nm2_Proj}
\end{figure}
%

What is the internal structure of this pair? In Fig.~\ref{fig:Nm2_Proj} the probability
of finding the second hole is shown when the first hole is projected to be at site 16,
namely at the center of the 32-sites chain used, and at orbital $a$. It is clear that
the probability for the second hole is the largest close to the first projected
hole, compatible with pairing. Moreover, the second hole is primarily at orbital $b$ if
the first is at orbital $a$. Thus, the pairs unveiled here involve holes primarily 
located at different orbitals. This will be shown below to be compatible with the pair-pair
correlation that is the most dominant in many 
portions of the phase diagram. Figure ~\ref{fig:Nm2_Proj}
also has prominent sharp peaks located at nearest-neighbors sites. Thus, the dominant hole
configuration in the pair is that of holes separated by just one lattice spacing, located
at different orbitals.

%
\begin{figure}[!ht]
\centering{
\subfloat{
\includegraphics[width=8.0cm, height=2.3cm, clip=true]
    {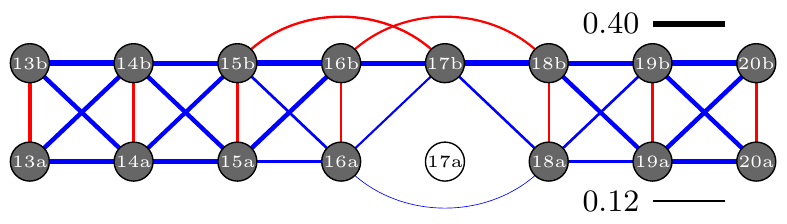}
    \llap{\parbox[b]{16cm}{\large\bf{(a)}\\\rule{0ex}{2.0cm}}}
}\\
\vspace{-0.4cm}
\subfloat{
\includegraphics[width=8.0cm, height=2.3cm, clip=true]
                {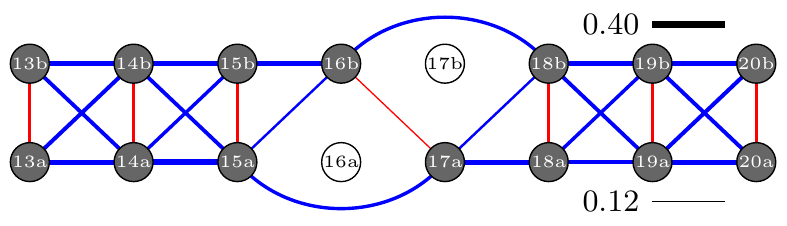}
    \llap{\parbox[b]{16cm}{\large\bf{(b)}\\\rule{0ex}{2.0cm}}}
}\\
\vspace{-0.4cm}
\subfloat{
\includegraphics[width=8.0cm, height=2.3cm, clip=true]
                {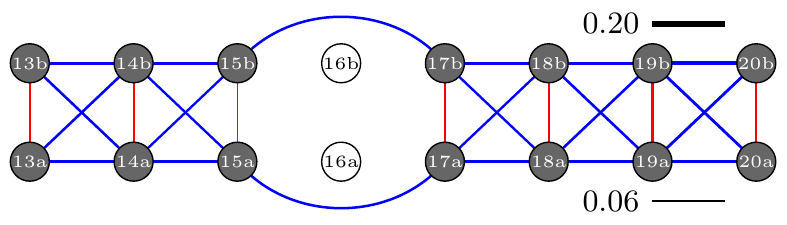}
    \llap{\parbox[b]{16cm}{\large\bf{(c)}\\\rule{0ex}{2.0cm}}}
}
}
\vspace{-0.5cm}
\caption{(color online) Structure of the spin-spin correlations for fixed projected 
arrangements of holes, using hole configurations with large weight in the ground states. 
(a) is the one-hole case; 
(b) corresponds to holes along the effective plaquette diagonals (largest weight for
two holes), while (c) are for holes along the effective rung.
All results are obtained using $32$ sites with two-orbitals at 
fixed $U/W = 1.60$ and $J_H/U = 0.25$. Blue lines correspond to AFM bonds 
while red lines are for FM bonds. Away from the holes, 
the expected pattern of FM effective rungs
and AFM legs is recovered. In the fixed hole configurations of panels 
(b) and (c), the normalized probability of the holes configuration is 
$\langle \mathcal{P}_{16a} \mathcal{P}_{17b} \rangle/\langle \mathcal{P}_{16a}\rangle = 0.169$ and
$\langle \mathcal{P}_{16a} \mathcal{P}_{16b} \rangle/\langle \mathcal{P}_{16a}\rangle = 0.0912$, 
respectively. In all cases there is a prominent ``across the hole'' AFM correlation. 
The case of two holes located along the same ``leg'' of the effective ladder has much
smaller weight in the two-hole ground state and it is not shown.
}
\label{fig:fig6_ProjSS}
\end{figure}
%

Projecting now one or two holes to particular locations and analyzing 
the spin-spin correlations in that framework leads 
to interesting conclusions. The results are shown in Fig.~\ref{fig:fig6_ProjSS}.
First, note that once the two-orbital chain results are displayed representing
each of the two orbitals by a chain, then
this illustrates that two-orbital chains can be mapped formally into a special case of 
one-orbital two-leg ladders. This is interesting in several respects, but here we wish to
emphasize the resemblance, once again, with the previously published 
results for two-orbital ladders~\cite{patel16}. 
Consider panel (a) for one hole: here the rungs of the effective ladder are ferromagnetic and the
legs are AFM. Thus, once the results are plotted as in Fig.~\ref{fig:fig6_ProjSS} the
magnetic order resembles the ``rung FM - leg AFM'' of BaFe$_2$S$_3$ as reported in~\cite{patel16}.
Also the AFM spin-spin correlation ``across the hole'' observed in early studies 
of models for cuprates~\cite{martins1,martins2,martins3} and
also found more recently in models for iron-based ladders~\cite{patel16} is present in panel (a).
From the spin perspective, ``across the hole'' AFM correlations 
are effectively equivalent to dropping sites of the chain, explaining 
the spin IC tendency in Fig.\ref{fig:Skx} with increasing doping. 

The results for two holes are equally interesting and also resemble those of previous investigations
for real iron-based ladder models. Panel (b) contains 
the hole arrangement with the largest probability in the two-holes ground state.
Similarly as in Fig.~10 of~\cite{patel16}, the plaquette diagonal opposite to the projected holes 
is FM and the ``across the hole'' antiferromagnetism is robust. 
Panel (c) shows the case where the two holes are along the rung (i.e. on-site in the real chain). 
Panels (b) and (c) are smoothly connected: for instance by moving the electron at ``17b'' to ``16b''
in panel (c), panel (b) is recovered if the spin correlations follow as if they were ``rubber bands''
attached to the electrons. Previous studies in models for cuprates have unveiled similar physics.

Ending this subsection, we will discuss a subtle effect related with the spin quantum number
of the two holes state. Employing Lanczos methods we studied the
total spin of the two-holes state employing both 
periodic and open boundary conditions (PBC and OBC, respectively) using
chains of length 4 and 6, at various Hund couplings and $U/W=1.60$. The behavior is erratic: while
for 6  sites and PBC the spin is always 0, for 4 sites with OBC it is always 1.
The interpretation of these results is difficult because of the 
 presence of the well-known edge states of Haldane chains when  OBC are used. 
Considering the difficulty in distinguishing between intrinsic 
spin quantum numbers of a hole pair vs. those at the edge of a Haldane chain, we do not
investigate further this topic and below we study both singlet and triplet pair channels to
find out which one dominates explicitly. The final result is that spin-singlet pairs are dominant
suggesting that the spin 1 quantum number found for some two-holes chains originates in the
edge states.

%
\begin{figure}[thbp]
\begin{center}
 \includegraphics[trim = 0mm 0mm 0mm 0mm,
 height=0.44\textwidth,width=0.47\textwidth,angle=0]{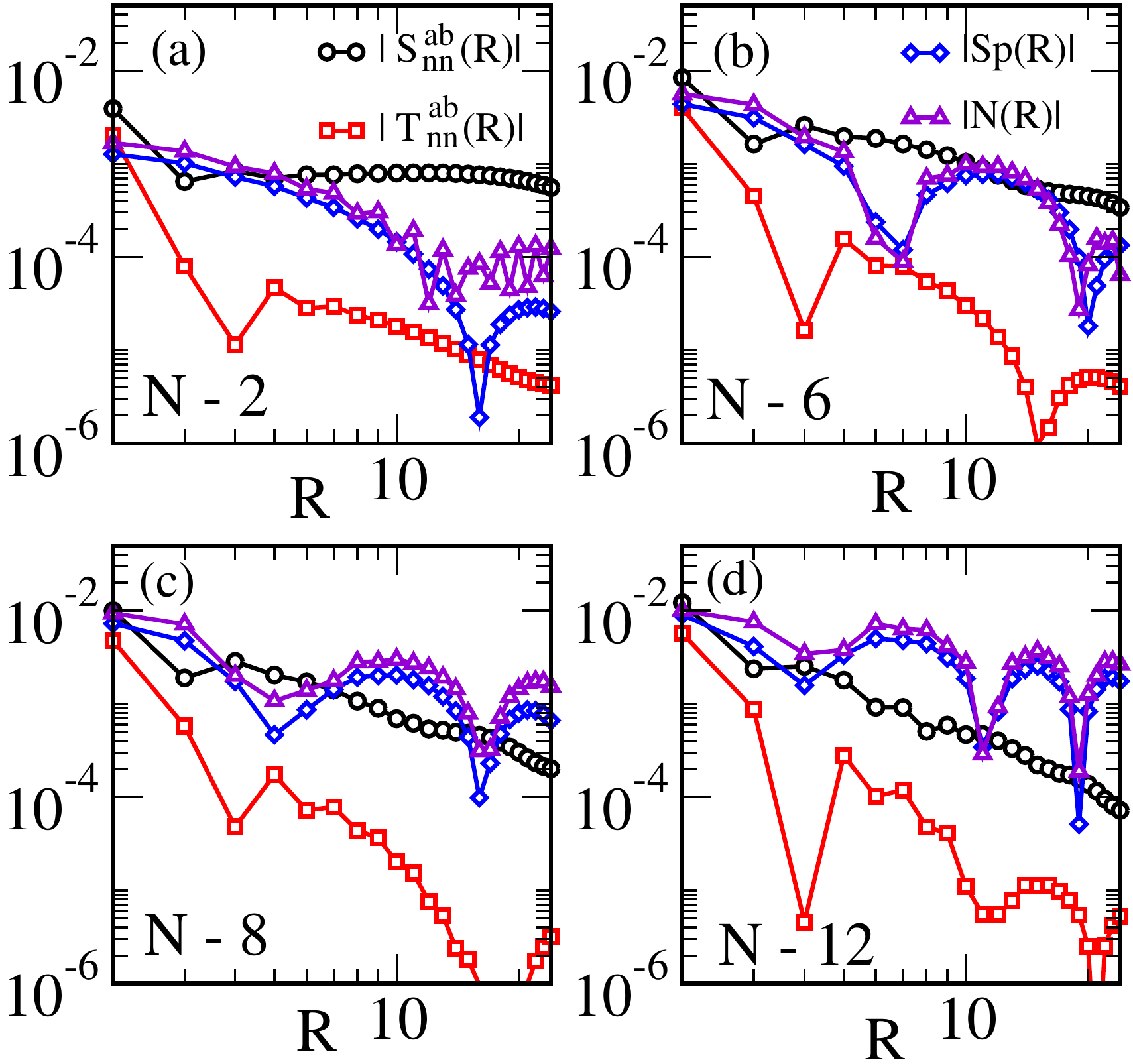}
\end{center}
\vspace{-0.6cm}
\caption{(color online) Real-space decay of the pair-pair (singlet and triplet), 
spin-spin, and charge-charge 
correlations involving
nearest-neighbor sites at fixed $U/W = 1.60$ and $J_H/U = 0.25$, and using a $48$-sites chain.
(a) corresponds to two holes doping, (b) to six holes doping, (c) to eight holes doping,
and (d) to twelve holes doping. 
}
\label{fig:Jhz_pair}
\end{figure}
%

\subsection{Pair-pair correlations and tendency to superconductivity}

The existence of hole binding at half-filling is often a precursor of superconducting
tendencies increasing doping. For this reason
we have measured the pair-pair correlations in all the channels described in Sec.~II, 
and contrasted their behavior with increasing distance 
against density-density and spin-spin correlations to find which channel dominates.
Representative 
results are shown in Fig.~\ref{fig:Jhz_pair}. Panel (a) contains results for 2 holes.
Here the pair-pair correlations are robust in the spin-singlet channel when involving different
orbitals and using nearest-neighbor sites, 
in agreement with the analysis of the internal structure of the pair in the previous subsection.
The analogous spin-triplet pair correlations decay much faster, while spin and charge correlations are
in between. However, in spite of the robustness of the singlet pair correlations in (a), 
the ground state only has two holes and these results, while promising, 
may be anomalous. More standard and exciting are the
results in panel (b) with 6 holes and a nominal hole doping $x=6/96 = 0.0625$ 
($N=96$ for a half-filled 48-sites cluster). The same spin-singlet
inter-orbital NN-sites pair correlation dominates here as well, as in (a). The decay with
distance is similar as in the charge and spin channels 
but only if the maxima is used for the latter. 
If, instead, the minima in charge and spin correlations are included 
in finding the most optimal fits then pair-pair correlations dominate. 
Note that the prominent oscillations in charge and spin correlations 
have been often reported before (for recent state-of-the-art efforts 
see~\cite{troyer}), although their origin is not fully clear; 
for the pair correlations a smoother behavior is often observed, as we found. 
As the number of holes increases, then the superconducting tendencies remain robust but 
diminish compared with spin and charge.
In panel (c) with $x=8/96$, the
pair-pair decay with distance approximately follows the average of spin and charge indicating
that they compete, while in panel (d) with $x=12/96$, pairing is already less 
robust than charge and spin channels. 
In summary, in a range of doping near half-filling and for the clusters that we studied, 
the superconducting correlations appear  to dominate, or
at the minimum decay at a similar rate as spin and charge. 
With increasing hole doping the importance of the 
pair-pair correlations diminishes.

%
\begin{figure}[thbp]
\begin{center}
 \includegraphics[trim = 0mm 0mm 0mm 0mm,
 height=0.34\textwidth,width=0.36\textwidth,angle=0]{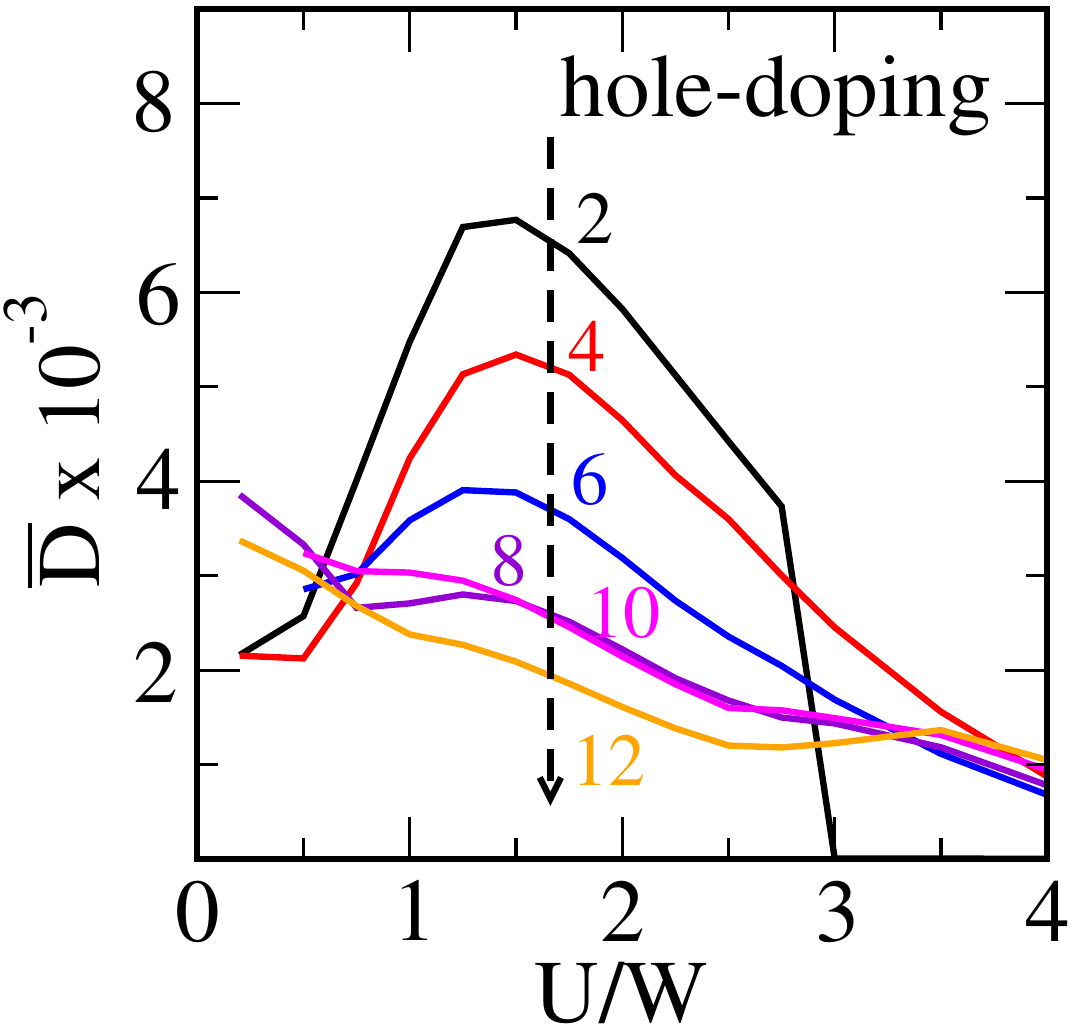}
\end{center}
\vspace{-0.6cm}
\caption{(color online) Pairing strength $\bar{D}$ (see Eq.~\ref{eq:pstregth}) vs. $U/W$, parametric
with number of holes. A 32-sites chain is used and $J_H/U = 0.25$.  
}
\label{fig:Dbar}
\end{figure}
%

In Fig.~\ref{fig:Dbar} we show the pairing strength $\bar{D}$, 
defined in Eq.~\ref{eq:pstregth},
as an indicator of the robustness of pairing correlations varying $U/W$ for various 
number of holes. Clearly it is the intermediate range of $U/W$ where
pairing dominates the most 
-- as found in the hole binding analysis -- and also when the numbers of holes is small.

%
\begin{figure}[thbp]
\begin{center}
 \includegraphics[trim = 0mm 0mm 0mm 0mm,
 height=0.45\textwidth,width=0.475\textwidth,angle=0]{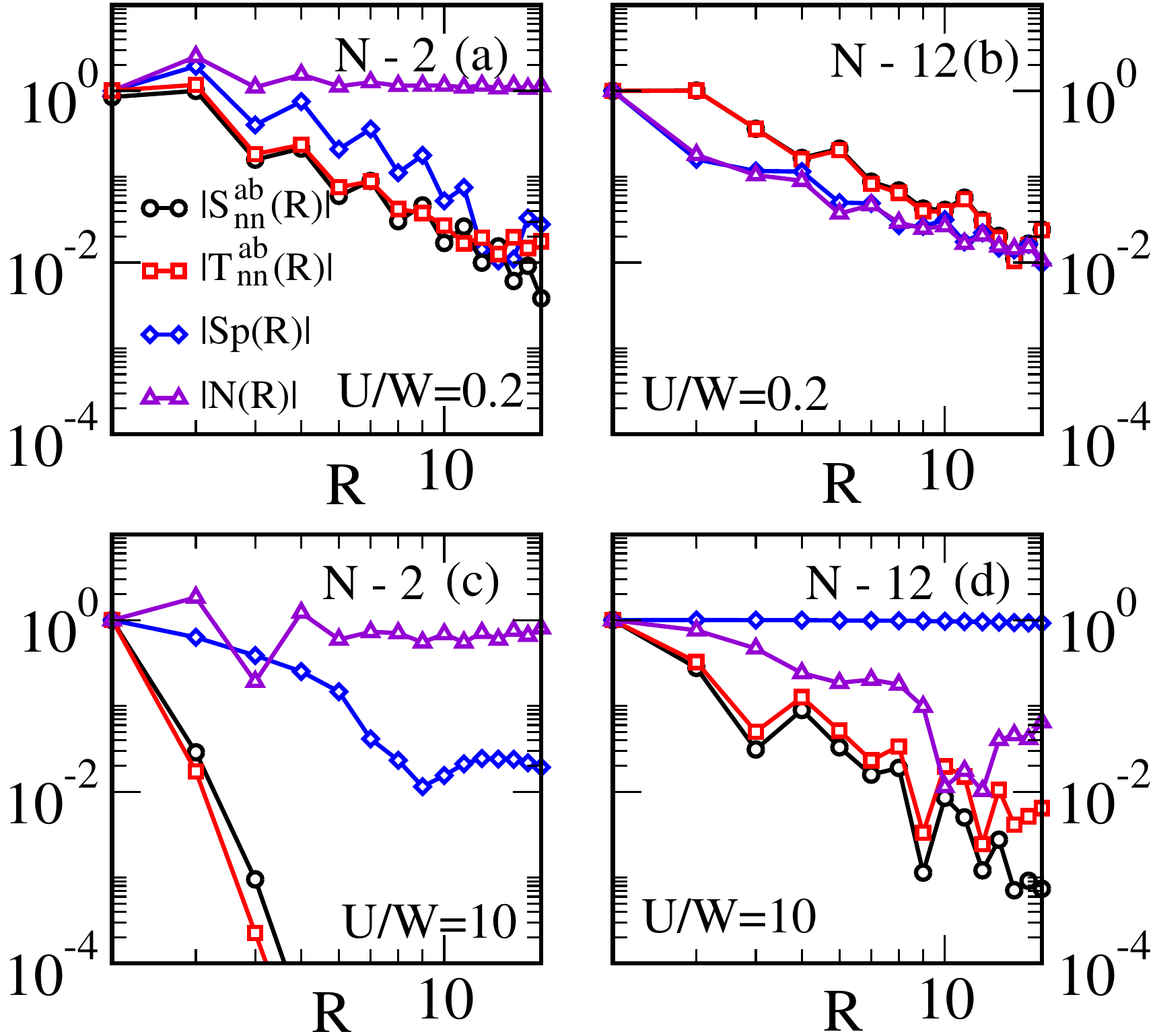}
\end{center}
\vspace{-0.6cm}
\caption{(color online) Real-space decay of the pair-pair (singlet and triplet), 
spin-spin, and charge-charge 
correlations for two and twelve holes at $U/W = 0.2$ and $10.0$, as indicated, 
and at $J_H/U = 0.25$ using a $32$-sites chain. 
In all cases, the decay of the pairs is either faster or approximately
equal to the spin and charge correlations. Thus, the pairing tendencies are robust
at intermediate coupling, compatible with the conclusions regarding hole binding.
}
\label{fig:ULargeSmall_Pair}
\end{figure}
%

Note that the presence of robust superconducting correlations in Fig.~\ref{fig:Jhz_pair}
occurs in the region of pair binding shown in Fig.~\ref{fig:fig1a_BindingvsU}.
Away from that region, for example
at small Hubbard coupling such as $U/W=0.2$ or at very large Hubbard coupling $U/W=10$
pairing is not as robust as at intermediate couplings, 
as illustrated in Fig.~\ref{fig:ULargeSmall_Pair}. Thus, once again we arrive to the 
conclusion that the behavior of the binding
energy and the pair correlation is compatible with one another.

For completeness, note that previous work also unveiled tendencies towards pairing in 
electronic two orbitals models
but under rather different circumstances. {\it (i)} For instance, in~\cite{riera97} Kondo models for
Y$_{2-x}$Ca$_x$BaNiO$_5$ were studied using Lanczos and DMRG techniques, supplemented
by AFM Heisenberg $J$ terms. The emphasis was on ferromagnetism and phase separation but
tendencies towards hole binding were also briefly reported. The signal for binding was strongest 
at high hole concentration such as $x=0.4$ and robust values of $J$ of order one. 
No pair-pair correlations were calculated, nor competition triplet vs. singlet was studied. 
{\it (ii)} In~\cite{shirakawa08} two one-orbital Hubbard
chains coupled by an explicitly ferromagnetic Heisenberg interaction were studied 
via bosonization and DMRG/Lanczos methods. Regions of singlet and triplet superconductivity were reported,
but note that this model has an explicit Heisenberg effective attraction, 
without a $U'$ repulsion (similar to our previous effort~\cite{xavierPRB10} to be discussed below).
The goal in~\cite{shirakawa08} was to study the singlet vs. triplet competition in superconductivity,
unlike our efforts that focus on unveiling pairing tendencies from a complete two-orbital Hubbard
model that is explicitly repulsive. {\it (iii)} In~\cite{ammon} a two-orbital Hubbard
model at $U=\infty$ was studied with emphasis on the influence of the Hund coupling. 
When $U'$ was included, charge-density waves were reported to dominate, while
in the absence of $U'$ but with robust $J_H$ then singlet or triplet pairing dominates. Our analysis,
on the other hand, focuses on a finite intermediate $U/W$ range where surprisingly
we found that singlet pairing dominates even in the presence of a realistic 
$U'>J_H$. As $U/W \rightarrow \infty$,
we found that hole binding no longer occurs, as shown in Fig.~\ref{fig:fig1a_BindingvsU}, 
and the charge or spin channels dominate over pairing 
(see panels (c,d) of Fig.~\ref{fig:ULargeSmall_Pair}) 
compatible with~\cite{ammon}. {\it (iv)} In~\cite{sano} results compatible
with ours were produced via the exact diagonalization
of a PBC 6-sites chain with emphasis on Luttinger liquid parameters 
using a two-orbital Hubbard model with a robust band splitting. 
{\it (v)} In~\cite{zegrodnik} using the statistically consistent Gutzwiller 
approximation for a square lattice, conclusions similar to ours 
were reached, reporting a stable spin-triplet $s$-wave superconducting 
state for a two orbital degenerate Hubbard model. This occurs, like 
in our case, even in the case $U' > J_H$ and near half-filling.

\FloatBarrier
\subsection{Role of Hund coupling and magnetic moments}

The origin of the pairing tendencies unveiled here is subtle and in this
subsection we report some observations to help clarify this matter. 
More work is needed to fully comprehend this hole pair formation, 
so ours are just the first steps in that direction.

One important factor correlated with the pairing we are reporting
is  the presence of well-formed magnetic moments
at every site. This is along the same direction as early studies of 
the $t-J$ model for cuprates~\cite{RMP94} where holes form bound states to reduce the damage
that mobile holes induce in an otherwise optimal antiferromagnetic arrangement. Each hole
alters the spin order 
in a finite region, and pairing of holes reduces the size of the regions
where spins are not properly arranged. This simple and well known notion must be at least part of the
explanation for our results because pairing in $\Delta E$, as shown in Fig.~\ref{fig:fig1a_BindingvsU},
occurs in regions where moments are well formed, as indicated in Fig.~\ref{fig:fig5_NMag}.

%
\begin{figure}[thbp]
\begin{center}
 \includegraphics[trim = 0mm 0mm 0mm 0mm,
 height=0.34\textwidth,width=0.36\textwidth,angle=0]{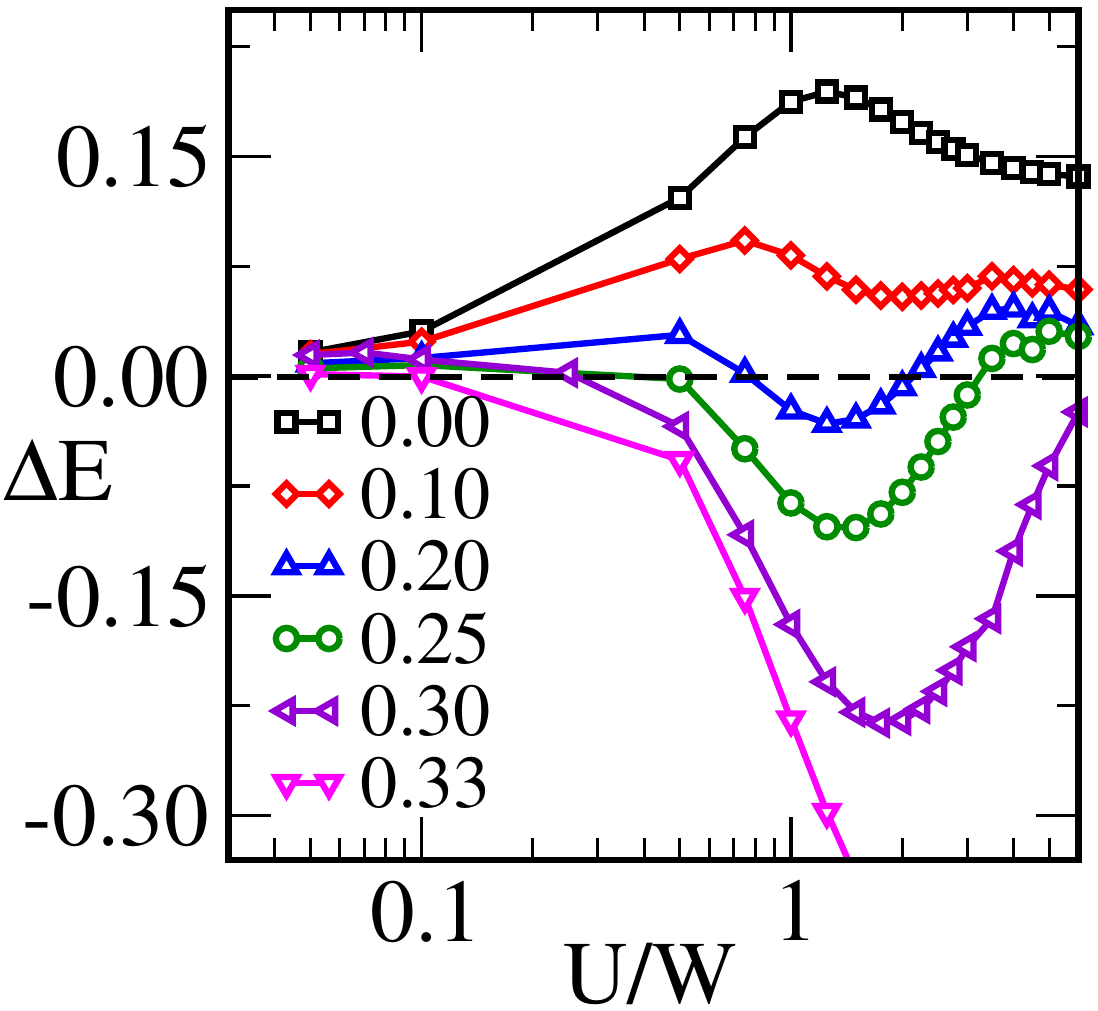}
\end{center}
\vspace{-0.6cm}
\caption{(color online)
$\Delta E$ vs. $U/W$ parametric with 
$J_H/U$ (inset) using an $8$-sites chain with two orbitals. }
\label{fig:fig4a_}
\end{figure}
%

In addition, we have observed that the Hund coupling in our model clearly is directly 
related to binding. Figure~\ref{fig:fig4a_} shows the binding energy in a wide range of $U/W$
parametric with $J_H/U$. At the smallest $J_H/U$ shown, the binding energy is positive and pairs
do not have a tendency to form. Consider now the special value $J_H/U=1/3$. In this case $J_H = U'$
because of the relation $U=U'+2J_H$. Thus, the natural repulsion $U'$ for two electrons at different
orbitals in the same lattice site is compensated by the natural tendency to bind induced by $J_H$.
In fact for $J_H/U=1/3$, and beyond i.e. $J_H/U > 1/3$, 
the binding energy $\Delta E$ is negative at all values of $U/W$. 

The reader should note 
that the connection between the realistic regime $J_H/U < 1/3$ and the unphysical
region $J_H/U > 1/3$ is non trivial. Naively, one may expect 
$\Delta E$ to be negative for all $U/W$ for
$J_H/U>1/3$, and positive for all $U/W$ for $J_H/U < 1/3$. However, the interpolation, while smooth, 
is more complex. Figure ~\ref{fig:fig4a_} shows that in the intermediate $U/W$ range, the binding
is negative for $J_H/U = 0.20$, $0.25$, and $0.30$, with a clear dip in the $U/W \sim 1-2$ range. This
dip, being smoothly connected with the broad negative binding energy region of $J_H/U = 1/3$, must be
caused by $J_H$ attraction effects that somehow are not fully compensated by $U'$ at intermediate couplings.

%
\begin{figure}[thbp]
\begin{center}
 \includegraphics[trim = 0mm 0mm 0mm 0mm,
 height=0.34\textwidth,width=0.36\textwidth,angle=0]{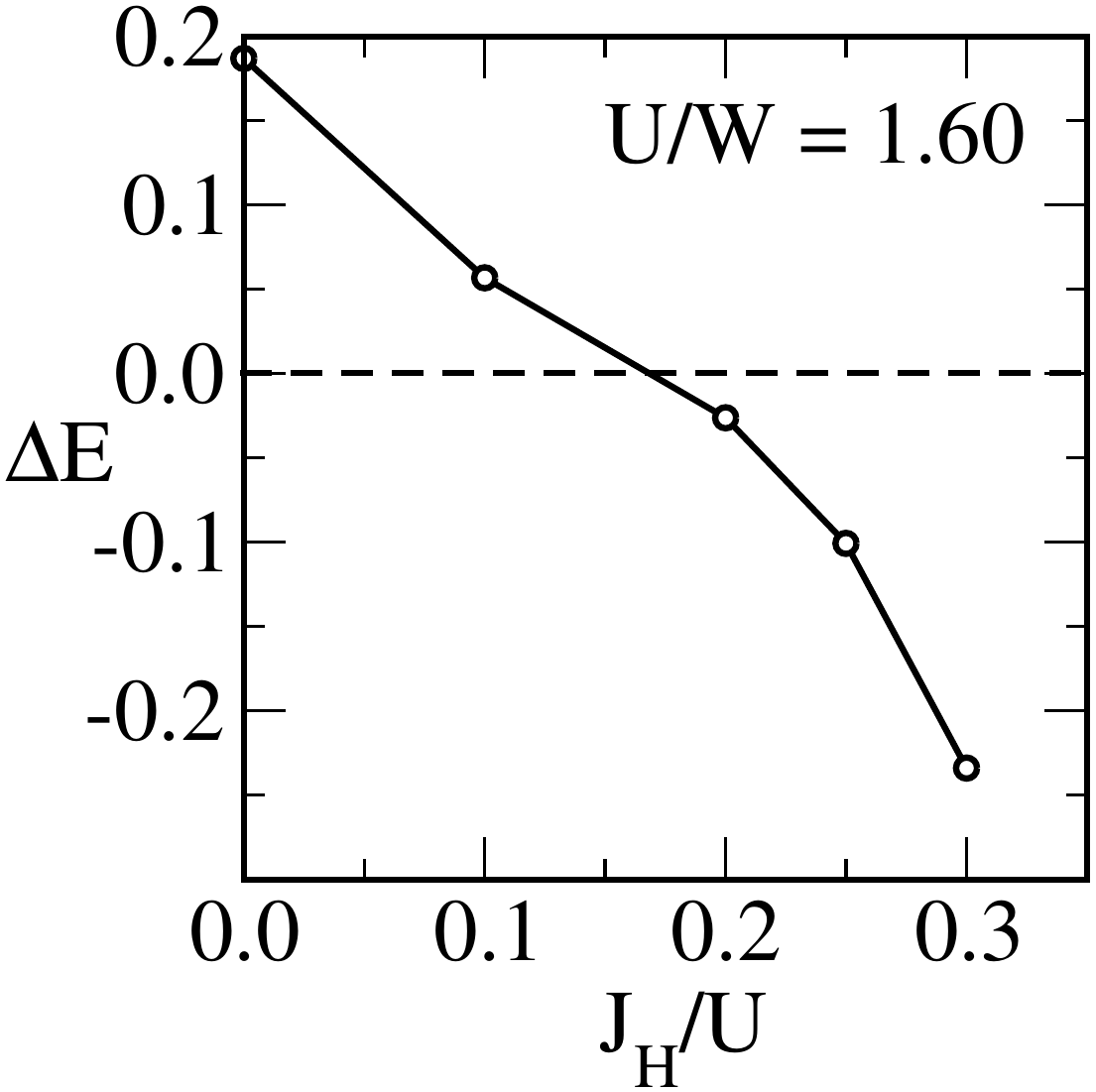}
\end{center}
\vspace{-0.6cm}
\caption{(color online) $\Delta E$ vs. $J_H/U$ at fixed $U/W = 1.60$ using
$16$ sites with two-orbitals. 
$\Delta E$ becomes significantly more negative as $J_H/U$ is increased.
}
\label{fig:fig4b_}
\end{figure}
%

The ``earlier'' than anticipated 
attractive effects of $J_H$ at intermediate $U/W$ 
are explicitly illustrated in Fig.~\ref{fig:fig4b_}
where $\Delta E$ is shown at the optimal $U/W = 1.60$ of our focus, varying $J_H/U$. At least for the small
system studied here, $\Delta E$ 
changes sign before $J_H/U=0.2$ and it becomes increasingly 
negative with further increasing $J_H$. 
While the causal effect of $J_H$ is clear, further work is needed to clarify how can this
attraction overcome the $U'$ repulsion in the intermediate coupling range. Moreover, the
attraction channel favors spin-singlets involving different orbitals at nearest-neighbor
sites. In fact, pairing in the spin-singlet inter-orbital NN-sites channels is enhanced
as $J_H$ increases as shown in panels (a,b) of Fig.~\ref{fig:JhHeis_pair}. 
Thus, it is a subtle combination of the Hund interaction together with 
antiferromagnetic short-range order that induces singlet 
pairing in this one-dimensional multiorbital model. More work is needed to clarify this interesting effect.

\subsection{Influence of additional inter-orbital Heisenberg interactions}

For completeness, we have also added an extra term to the Hamiltonian in order
to boost pairing tendencies. This term is simply a Heisenberg spin-spin interaction
with strength $J_D$ defined as

%
\begin{equation} \label{eq:Heis}
\begin{split}
H_{D} &= J_{D} \sum_{\substack{\langle i j \rangle}} 
{{\mathbf{S}_{i a}}\cdot{\mathbf{S}_{j b}}}.
\end{split}
\end{equation}
%

\begin{figure}[thbp]
\begin{center}
 \includegraphics[trim = 0mm 0mm 0mm 0mm,
 height=0.41\textwidth,width=0.49\textwidth,angle=0]{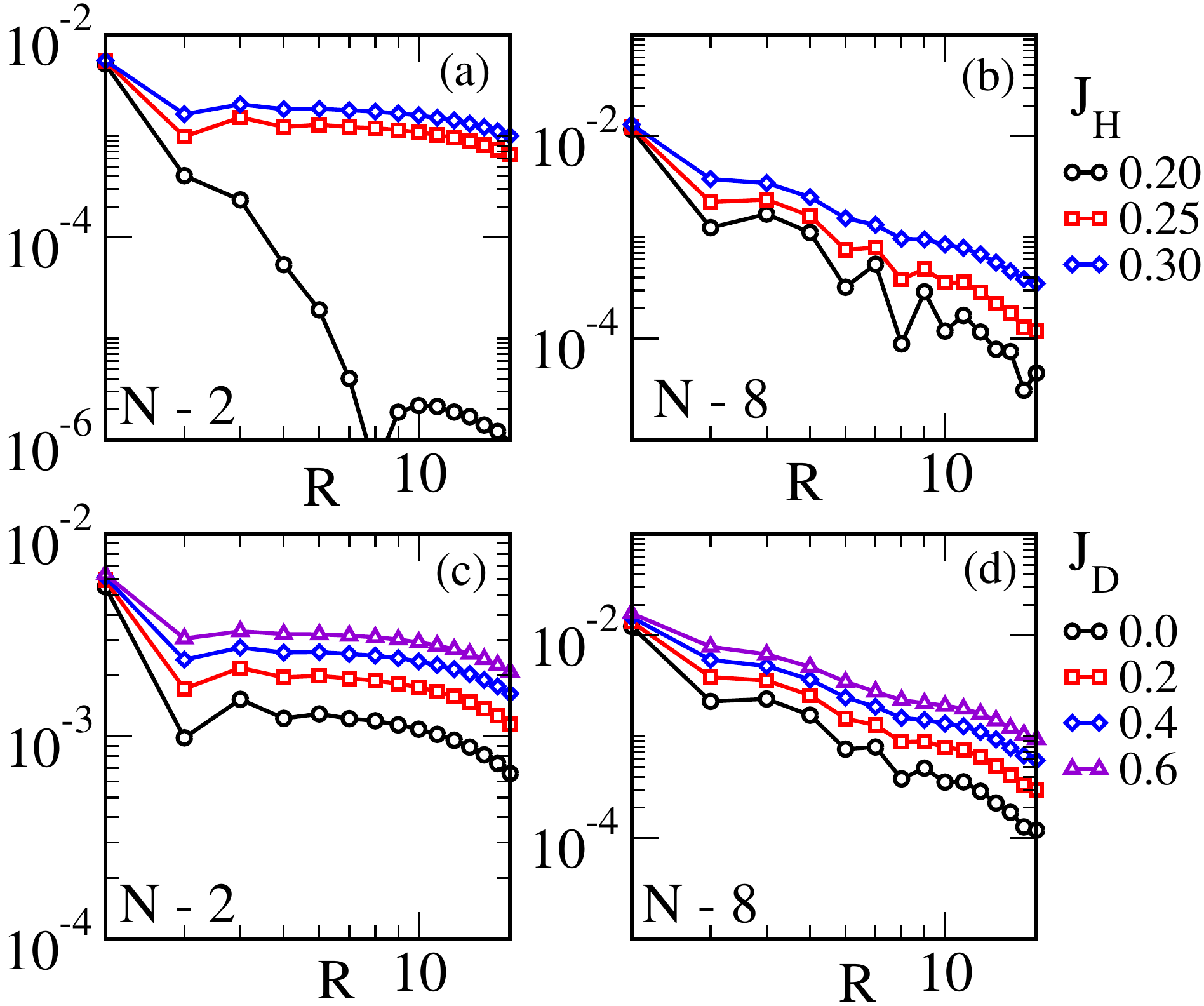}
\end{center}
\vspace{-0.6cm}
\caption{(color online) Decay of pair-pair singlet correlations $S^{ab}_{nn}$ 
at $U/W = 1.60$, for various Hund couplings in panels (a) and (b), and for various diagonal 
AFM Heisenberg couplings $J_D$ in panels 
(c) and (d). Shown are results for 2 and 8 holes, as indicated, using in all
cases a 32-sites chain and $J_H/U = 0.25$. 
Clearly increasing both $J_H$ and $J_D$ magnify the pairing tendencies.
}
\label{fig:JhHeis_pair}
\end{figure}

\noindent The motivation for adding this term is two folded. First, it plays a role
similar to that of ``$J$'' in the standard $t-J$ model, and we know that increasing $J$
increases pairing tendencies~\cite{RMP94}. Second, the new term links 
the spins of two electrons located at NN sites and
at different orbitals (note orbital indexes in Eq.~\ref{eq:Heis}), 
resembling the structure of the pairs that we have found above. In agreement
with these expectations indeed we have observed that pairing tendencies in the
dominant spin-singlet NN-sites inter-orbital channel are enhanced as shown in panels
(c) and (d) of Fig.~\ref{fig:JhHeis_pair}. A similar analysis adding instead
a NN Heisenberg coupling between electrons in the $same$ orbital only
showed minor changes in the decay of the correlations (not shown). 
%
%
Clearly the electrons/holes 
in the dominant pairs have a preference to be in different orbitals.
Future studies of superconductivity in the two-orbital Hubbard model analyzed here 
can benefit from enhanced pairing effects by including $J_D$.

\FloatBarrier
\section{Conclusions}

In this publication we have investigated the magnetic, hole pairing, and superconducting
properties of a two-orbital Hubbard model defined on a chain. The primary motivation
is the recent report of hole binding tendencies in a similar model but defined on a 
two-leg ladder geometry~\cite{patel16}, motivated by the discovery of superconductivity under high pressure 
in the ladder compound BaFe$_2$S$_3$~\cite{NatMatSC}. In that previous computational effort, 
the binding tendency was found to be negative, thus signaling pairing, but the results 
could not be confirmed beyond small systems 2$\times$8. In addition 
pair-pair correlations were not measured in that early effort. In the
present work much longer chains can be studied and a variety of correlation functions 
were measured and their decay with distance compared to decide which is dominant.
In the same spirit as in \cite{patel16}, here our search for superconducting tendencies was 
based on hole doping while the experimental setup relied on pressure. The 
$ab$-$initio$ calculations in \cite{dong17} justify our theoretical approach because 
they reported that pressure leads to modifications in the average electronic 
density at the iron atoms.

Our results are interesting for several reasons. The data reported here for the binding
energy resemble those of the ladder, but on chain sizes up to $64$ sites. Size scaling shows
that the results survive the bulk limit. Qualitatively both for ladders and chains it is
the intermediate region of $U/W$ were binding does occur. Having 
almost saturated magnetic moments is important together 
with a robust Hund coupling. Neither
very weak nor very strong $U/W$ coupling seem suitable for 
pairing, a conceptually interesting result. The absence of pairing at very large $U/W$ may be
related with competing ferromagnetic tendencies when holes are added, as in double exchange models.
This line of research is being investigated at present.

Moreover, by measuring pair-pair correlations in the spin singlet channel, 
and using pair operators involving different orbitals and nearest-neighbor sites, 
a region of hole density and couplings was identified where
superconducting correlations decay slower, or 
at least at the same rate, than spin and charge correlations.
Having different orbitals and nearest-neighbor 
sites is compatible with the internal structure of the pair. 

By varying the Hund coupling into the region believed to be unphysical 
where $J_H$ becomes as large as $U'$ (this occurs at $J_H/U=1/3$ 
if the standard relation $U=U'+2J_H$
is assumed~\cite{xavierPRB10,xavierPRL}), 
then an unexpected smooth continuity was observed between $J_H/U > 1/3$, where binding
occurs at all values of $U/W$ because $J_H$ becomes an effective attraction when it 
overcomes $U'$, 
and the region widely believed to be realistic $J_H/U \sim 0.25$. 
This smooth continuity occurs primarily at
intermediate $U/W$ couplings. Thus, for reasons that still 
need better clarification the effective $J_H$
attraction at $J_H/U > 1/3$ can become operative 
even at smaller Hund couplings in a reduced $U/W$ range.
The chosen dominant channel involves holes in different 
orbitals, a spin-singlet combination, and nearest-neighbors sites.

The observation that pairing, charge, and 
spin correlations are sometimes of similar strength, 
as in panel (c) of Fig.~\ref{fig:Jhz_pair} for $N-8$ electrons (48 sites), 
suggests that future work should also address the possible formation of
``pair density waves''. These are subtle broken-symmetry states that intertwine charge density waves,
spin density waves, and superconducting orders. In this state the superconducting order parameter is
spatially modulated in such a way that the uniform component is zero or very small, but it has a strong
oscillatory component~\cite{berg,almeida,fradkin}.

In summary,  these results contribute towards understanding pairing tendencies
in quasi one-dimensional iron-based superconductors.
Binding was found to occur at intermediate couplings, a regime that previous studies showed 
to be realistic for chalcogenides~\cite{Dainature,RMP}. 
There is plenty of work ahead. While superconducting 
correlations already appear to dominate at low hole doping, these
results must be confirmed using even longer chains. Moreover, 
although it seems clear that a robust Hund coupling
and robust magnetic moments are needed, developing an 
even more detailed qualitative understanding of the origin of
pairing is important. Our group will continue working 
along these  lines in the near future.

\FloatBarrier
\section*{Acknowledgments}
N.D.P., A.M., and E.D. were supported by the National
Science Foundation Grant No. DMR-1404375. A.N. was 
supported by the U.S. Department of Energy (DOE),
Office of Basic Energy Science (BES), Materials Science and
Engineering Division. Part of this work was conducted at the
Center for Nanophase Materials Sciences, sponsored by the
Scientific User Facilities Division (SUFD), BES, DOE, under
contract with UT-Battelle. G.A. acknowledge support
by the Early Career Research program, SUFD, BES, DOE.
Computer time was provided in part by resources supported
by the University of Tennessee and Oak Ridge National
Laboratory Joint Institute for Computational Sciences.

\end{document}